\documentclass[prl,showpacs,amsmath,amssymb]{revtex4-1}
\usepackage{graphicx}
\usepackage{dcolumn}
\usepackage{bm}
\usepackage{amsmath}
\usepackage{amssymb}
\usepackage[usenames]{color}
\usepackage[normalem]{ulem}
\usepackage[none]{hyphenat}

\begin{document}

\title {Non-Dirac topological surface states in (SnTe)$_{n\geq2}$(Bi$_2$Te$_3$)$_{m=1}$}

\author{S.~V. Eremeev}
 \affiliation{Institute of Strength Physics and Materials Science,
634021, Tomsk, Russia}
 \affiliation{Tomsk State University, 634050 Tomsk, Russia}
 \affiliation{Saint Petersburg State University, Saint Petersburg, 198504,
 Russia}

\author{T.~V. Menshchikova}
 \affiliation{Tomsk State University, 634050 Tomsk, Russia}

\author{I.~V. Silkin}
 \affiliation{Tomsk State University, 634050 Tomsk, Russia}

\author{M.~G. Vergniory}
\affiliation{Donostia International Physics Center (DIPC),
             20018 San Sebasti\'an/Donostia, Basque Country,
             Spain\\}

\author{P.~M. Echenique}
 \affiliation{Donostia International Physics Center (DIPC), 20018 San Sebasti\'an/Donostia, Basque Country, Spain\\}
\affiliation{Departamento de F\'{\i}sica de Materiales UPV/EHU,
Centro de F\'{\i}sica de Materiales CFM - MPC and Centro Mixto
CSIC-UPV/EHU, 20080 San Sebasti\'an/Donostia, Basque Country, Spain}

\author{E.~V. Chulkov}
\affiliation{Donostia International Physics Center (DIPC),
             20018 San Sebasti\'an/Donostia, Basque Country,
             Spain\\}
\affiliation{Departamento de F\'{\i}sica de Materiales UPV/EHU,
Centro de F\'{\i}sica de Materiales CFM - MPC and Centro Mixto
CSIC-UPV/EHU, 20080 San Sebasti\'an/Donostia, Basque Country, Spain}
 \affiliation{Tomsk State University, 634050 Tomsk, Russia}
 \affiliation{Saint Petersburg State University, Saint Petersburg, 198504,
 Russia}


\begin{abstract}

A new type of topological spin-helical surface states was discovered
in layered van der Waals bonded
(SnTe)$_{n=2,3}$(Bi$_2$Te$_3$)$_{m=1}$ compounds which comprise two
covalently bonded band inverted subsystems, SnTe and Bi$_2$Te$_3$,
within a building block. This novel topological states demonstrate
non-Dirac dispersion within the band gap. The dispersion of the
surface state has two linear sections of different slope with
shoulder feature between them. Such a dispersion of the topological
surface state enables effective switch of the velocity of
topological carriers by means of applying an external electric
field.

\end{abstract}


\maketitle

\section{Introduction}

A topological insulators (TIs) are materials with a gap in bulk
electronic spectrum driven by the spin-orbit interaction (SOI),
which is topologically distinct from the ordinary insulators. This
distinction, characterized by a $\mathbb{Z}_2$ topological
invariant, is manifested in the existence of the
time-reversal-symmetry protected spin-helical gapless surface state
having linear (or almost linear) Dirac dispersion.

Starting from the discovery of the TI phase in the quintuple layer
(QL) structured binary compounds Bi$_2$Te$_3$, Bi$_2$Se$_3$, and
Sb$_2$Te$_3$ and its thorough experimental and theoretical
investigation
\cite{Xia_NatPhys2009,Chen_Sci2009,Zhang_PRL2009,Kuroda_PRL2010,Zhang_NatPhys2009,Zhang_NatPhys2010,Zhang_NJP2010,Eremeev_JETPL,Eto_PRB2010}
a huge number of $\mathbb{Z}_2$ TIs were found. The most of known
TIs are Bi$_2$Te$_3$-type ternary and quaternary layered materials.
Among them QL and septuple layer (SL) structured compounds as well
as the materials with alternating QL and SL units were theoretically
predicted and experimentally
confirmed.\cite{Eremeev_NatComm,Kuroda_PRL2012,Neupane_PRB2012,Okuda_PRL2013}

Another class of topological materials is materials with band gap
inverted by strong spin-orbit interaction in which the topological
surface states are protected by the crystal mirror symmetry
\cite{Fu_TCI} -- so-called topological crystalline insulators (TCI).
One representative of a TCI is SnTe, \cite{Hsieh,Tanaka} showing an
even number of the Dirac cones in the surface spectrum.

Recently it was demonstrated that in the Bi$_2$Te$_3$/SnTe
heterostructure, the system containing $\mathbb{Z}_2$ and crystal
mirror symmetry TIs, the $\bar\Gamma$ Dirac surface states of
Bi$_2$Te$_3$ and SnTe annihilate at the interface while the Dirac
state of SnTe at the $\bar {\rm M}$ ``survives".\cite{Rauch} This
points out that the interaction of $\mathbb{Z}_2$ TI and TCI systems
can results in unusual change in topological surface states picture.

\begin{figure}
\includegraphics[width=0.5\columnwidth]{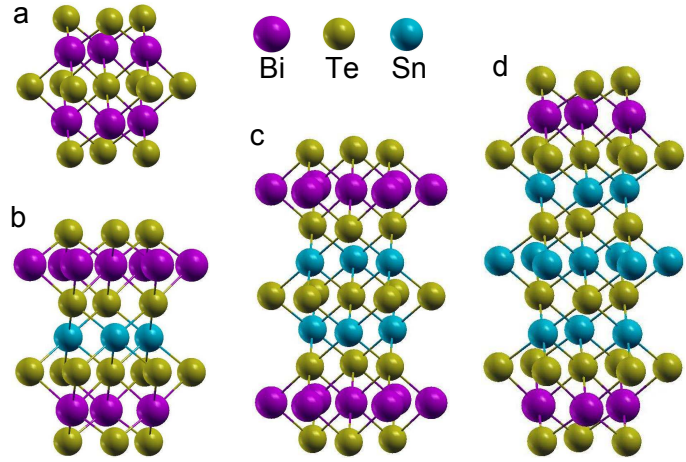}
\caption{(Color online) Building blocks of Bi$_2$Te$_3$ (a);
SnBi$_2$Te$_4$ (b); Sn$_2$Bi$_2$Te$_5$ (c); and Sn$_3$Bi$_2$Te$_6$
(d).}
 \label{fig1}
\end{figure}

Different systems, containing layers of band-inverted Bi$_2$Te$_3$
and SnTe materials are (SnTe)$_{n}$(Bi$_2$Te$_3$)$_m$
compounds.\cite{Kuropatwa} In this series the $(n=1, m>1)$ compounds
-- SnBi$_4$Te$_7$, SnBi$_6$Te$_{10}$, and SnBi$_8$Te$_{13}$ are
structurally identical to earlier studied lead-based TIs
\cite{Eremeev_NatComm}, containing alternating QL and SL structural
units, while the $(n\geq1, m=1)$ compounds are formed by unique,
nonalternating building blocks. The simplest system $(n=1, m=1)$,
i.e. SnBi$_2$Te$_4$ is composed by SL hexagonally ordered blocks
stacked along the $c$ axis and separated by van der Waals spacings.
The SL building block (Fig.~\ref{fig1}(b)) can be obtained from the
original QL block of Bi$_2$Te$_3$ (Fig.~\ref{fig1}(a)) by
introducing the SnTe bilayer between the Bi atomic layer and the
central Te one. This compound was predicted to be 3D $\mathbb{Z}_2$
TI with single Dirac cone in the bulk energy gap
\cite{Eremeev_NatComm}. The Sn$_2$Bi$_2$Te$_5$ compound $(n=2, m=1)$
is formed by nonuple layer (NL) building blocks \cite{Kuropatwa}.
The NL block (Fig.~\ref{fig1}(c)) is constructed from SL block by
introduction additional SnTe hexagonal bilayer in the middle of the
SL or two bilayers into original QL. Thus this compound differs from
Bi$_2$Te$_3$/SnTe heterostructure having the ionic-covalent bonded
Bi$_2$Te$_3$ and SnTe atomic layers. Another experimentally
confirmed structure (SnTe)$_{n}$(Bi$_2$Te$_3$)$_m$, which is also
built from the ionic-covalent bonded Bi$_2$Te$_3$ and SnTe atomic
layers is Sn$_3$Bi$_2$Te$_6$ $(n=3, m=1)$.\cite{Kuropatwa} In this
case additional SnTe layers are incorporated into the building block
(Fig.~\ref{fig1}(d)). We are not aware of the possibility of growth
of stable (SnTe)$_{n}$(Bi$_2$Te$_3$)$_{m=1}$ compounds with larger
$n$ which have building blocks of the same type.

In the present paper we demonstrate by means of relativistic density
functional theory (DFT) calculations that
(SnTe)$_{n=2,3}$(Bi$_2$Te$_3$)$_{m=1}$ compounds are $\mathbb{Z}_2$
TIs, in which owing to complex band inversion of bulk states
belonging to Bi$_2$Te$_3$ and SnTe subsystems at the surface arise
the topological surface state (TSS) demonstrating non-Dirac, dog-leg
dispersion within the bulk energy gap. We discuss the localization
and spin texture of the TSS in the context of gate control of the
topological transport properties. We also demonstrate the existence
of such a TSS in Sn$_2$Sb$_2$Te$_5$ compound.

For calculations we use the Vienna Ab Initio Simulation Package
(VASP) \cite{VASP1,VASP2} with generalized gradient approximation
(GGA) \cite{PBE} for the exchange correlation potential. The
interaction between the ion cores and valence electrons was
described by the projector augmented-wave method.\cite{PAW1,PAW2}
Relativistic effects, including SOI, were taken into account. The
atomic positions of bulk (SnTe)$_{n=2,3}$(Bi$_2$Te$_3$) compounds
were optimized. In slab calculations a vacuum space of $\sim 20$
\AA\ was included to ensure negligible interaction between opposite
surfaces. Complementary calculations for bulk band structure were
performed using the fullpotential linearized augmented plane-wave
method as implemented in the FLEUR code.\cite{FLEUR}

\begin{figure}
\includegraphics[width=0.5\columnwidth]{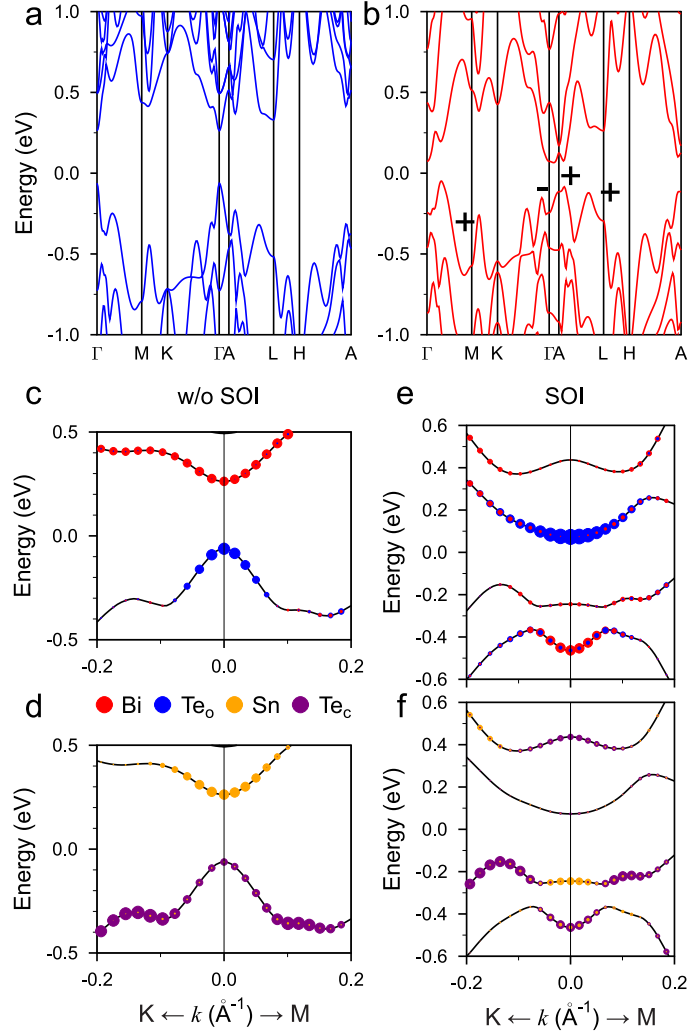}
\caption{(Color online) Bulk electronic structure of
Sn$_2$Bi$_2$Te$_5$ calculated without (a) and with (b) SOI included
(Signs of band parities at the TRIM are also shown). Atomic
composition of the near-gap states in bulk Sn$_2$Bi$_2$Te$_5$
calculated without (c,d) and with (e,f) SOI included.
 }
 \label{fig2}
\end{figure}

\begin{figure*}
\includegraphics[width=\textwidth]{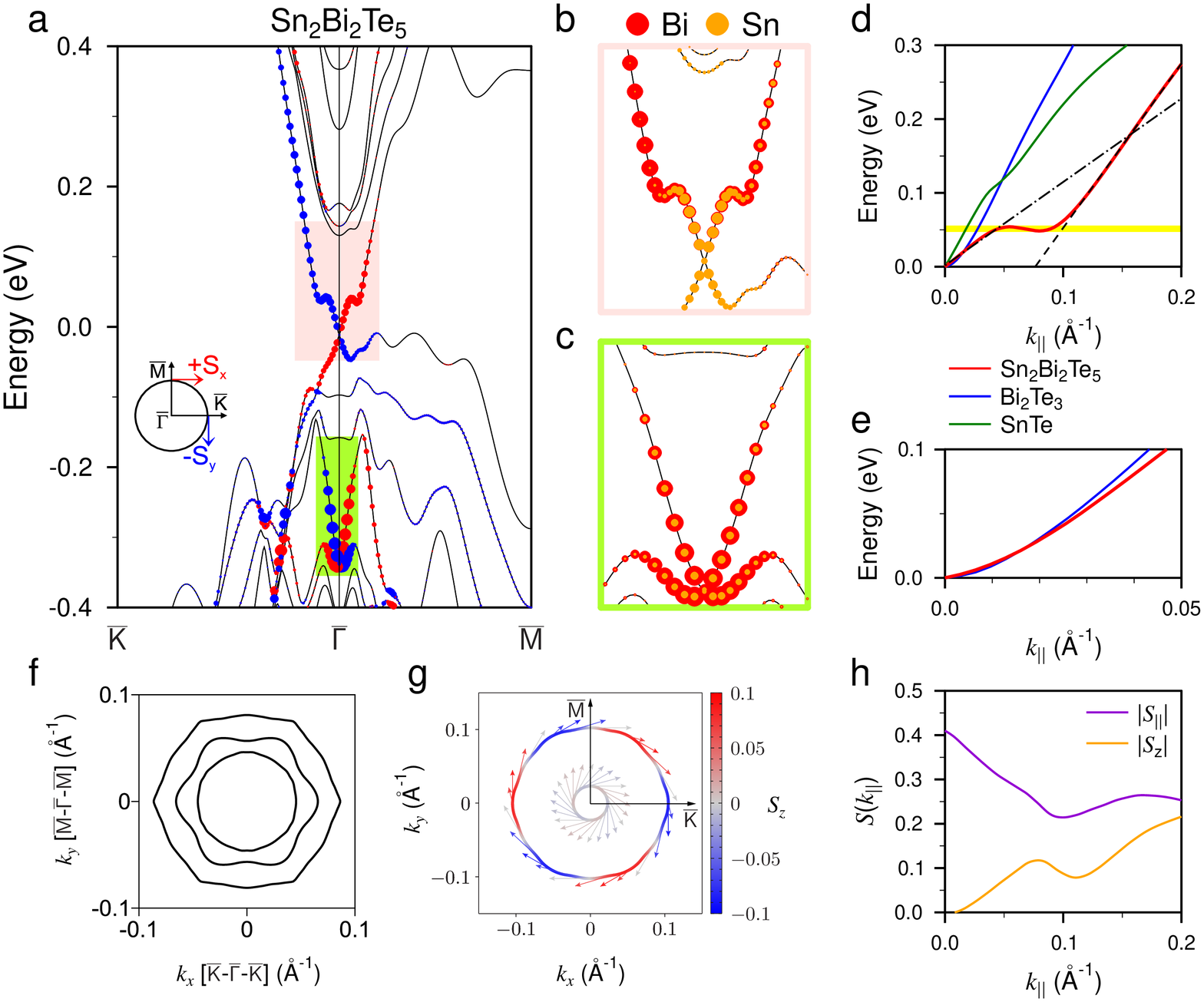}
\caption{(Color online) (a) Surface electronic structure of
Sn$_2$Bi$_2$Te$_5$; The size of the color circles represent the
weight of the states in the outermost building block (NL) of the
slab; red and blue colors show the sign of the in-plane ($S_x$ and
$S_y$) spin components; lightpink and lightgreen rectangles mark the
areas of the bandgap and valenceband topological surface states,
respectively. Atomic weights of Bi and Sn in the bandgap (b) and
valenceband (c) TSSs. (d) Energy dispersion of the bandgap TSS
compared to Bi$_2$Te$_3$ and SnTe $\bar\Gamma$ Dirac cones. Dashed
and dot-dashed lines are linear fit for the TSS above and below
shoulder, respectively. (e) Dispersion of the valenceband TSS
compared to Bi$_2$Te$_3$. (f) Constant energy contours in the
shoulder region (yellow stripe in panel (d)). (g) Spin-resolved
constant energy contours for bandgap TSS at energies of 25 meV below
and above the shoulders. Arrows adjacent to the contours denote the
in-plane spin component $S_\|$. The out-of-plane spin component
$S_z$ is indicated by the color with red and blue corresponding to
the positive and negative values, respectively. (h) The spin
components $S_\|$ and $S_z$ for bandgap TSS as functions of $k_\|$
along the $\bar\Gamma$-$\bar{\rm K}$ direction.}
 \label{fig3}
\end{figure*}

\section{Results}

The bulk band structures of Sn$_2$Bi$_2$Te$_5$ shown in
Fig.~\ref{fig2} have been calculated with and without SOI included.
As can be seen from Fig.~\ref{fig2}(a), the direct energy gap of
0.32 eV exist at the $\Gamma$ point in the calculation without SOI
which transforms into indirect gap of 0.12 eV under switch on the
SOI (Fig.~\ref{fig2}(b)). The time-reversal invariant momenta
(TRIMs) in this Brillouin zone are the $\Gamma$, A and triply
degenerate M and L points. We checked the parity of the
wavefunctions at all TRIMs to determine the topological character of
this material and found that the parity inversion of bulk bands
occurs at the $\Gamma$ point which leads to $\mathbb Z_2$ invariants
1;(000) like in Bi$_2$Te$_3$.

In spite of this similarity in SOI-induced wavefunction parity
change, the total band inversion at $\Gamma$ is an entanglement of
two different inversions corresponding to each subsystem due to the
complex crystal structure of the compound. For instance, in
Bi$_2$Te$_3$ the SOI modifies the top valence band state
(predominantly of $p_z$ character) of Te atoms lying in the outer
layers of QL (Te$_{\rm o}$) and the Bi $p_z$-state of the bottom
conduction band so that they are inverted in the close vicinity of
the $\Gamma$ point. A similar simple band inversion occurs in the
bulk SnTe where Sn and Te states are inverted in the L point of the
cubic Brillouin zone. Instead, in Sn$_2$Bi$_2$Te$_5$ the spectra
calculated without SOI both Bi and Sn states form the bottom of the
conduction band while the states of Te$_{\rm o}$ atoms and tellurium
atoms of the central layer of NL (Te$_{\rm c}$) lie in the top
valence band (Fig.~\ref{fig2}(c,d)). Switching on the SOI results in
an inversion of Bi and Te$_{\rm o}$ states and Sn and Te$_{\rm c}$
states. However, in contrast to the binary systems this inversion
has more complex character. As can be seen in Fig.~\ref{fig2}(e)
after switching on the SOI the Te$_{\rm o}$ states become dominant
in the conduction band while the Bi states (in contrast to the
Bi$_2$Te$_3$ case) occupy the second valence band in the vicinity of
$\Gamma$. The SOI-inverted Sn states lie in the highest valence band
and Te$_{\rm c}$ states appear in the second conduction band
Fig.~\ref{fig2}(f). Additionally, the SOI-inverted Sn states (with
smaller weight) appear in the second valence band. Thus, the
Sn$_2$Bi$_2$Te$_5$ TI has a competitive bulk $\Gamma$-gap band
inversion which comprises the inversion of the states belonging to
different subsystems, SnTe and Bi$_2$Te$_3$, so that the valence
band edge is mainly formed by Sn states and the conduction band edge
occupied by Te$_{\rm o}$ states while the Bi and Te$_{\rm c}$ states
lie next to the edges bands.

Such a superposition of the SOI-induced bulk band inversions in the
the SnTe and Bi$_2$Te$_3$ subsystems of the
(SnTe)$_2$(Bi$_2$Te$_3$)$_1$ TI results in formation of unusual
topological states on the surface. In the surface spectrum two
topological surface states arise at the $\bar\Gamma$ point. The
upper TSS lies in the band gap (see lightpink rectangle in
Fig.~\ref{fig3}a) while the lower TSS (lightgreen rectangle in
Fig.~\ref{fig3}a) resides in the valence band (bandgap and
valenceband TSS's, respectively from now on). The lower, valenceband
TSS, propagating within the local valence band gap, has typical
Dirac dispersion for TIs. In contrast, the bandgap TSS displays
nonlinear (non-Dirac) dog-leg behavior. It has two linear sections
in the spectrum: near the band crossing (Dirac point) and below the
conduction band which connect to each other in the middle part of
the gap with formation of shoulders in the spectrum. Like in other
TIs, the TSSs in Sn$_2$Bi$_2$Te$_5$ are localized within outermost
building block of the slab. However, in this case, where the
building block has two band-inverted SnTe and Bi$_2$Te$_3$
subsystems, the localization of the TSS is affected by the
peculiarities in the bulk band inversion. Figs.~\ref{fig3}(b,c) show
the Bi and Sn atomic weights in the TSSs. As one can see in
Fig.~\ref{fig3}(c), the valenceband TSS has dominant Bi weights that
reflect its dominant localization in the outer Bi-Te layers of NL.
Comparing the dispersion of this state with that of the Dirac cone
in Bi$_2$Te$_3$\cite{Eremeev_JETPL} (Fig.~\ref{fig3}(e)) one can
conclude that at small $k_\|$ (where the valenceband TSS has
localized character) both bands are almost identical. In the bandgap
TSS (Fig.~\ref{fig3}(b)), the lower linear section near the band
crossing, the Sn contribution dominates and thus the state is mainly
localized in the inner Sn-Te layers of NL. Approaching the shoulders
the Sn contribution disappears rather quickly and above the band
shoulders the TSS becomes to be mostly localized in the outer Bi-Te
layers. Thus the band shoulders in the TSS spectrum are associated
with the change of the TSS localization from the SnTe to
Bi$_2$Te$_3$ subsystem. It should be noted that the slope of the
upper linear section of the bandgap TSS (at $k_\|>0.1$ \AA$^{-1}$)
localized in the outer Bi-Te layers is comparable with the slope of
the Bi$_2$Te$_3$ Dirac cone (Fig.~\ref{fig3}(d)), while the
dispersion of the lower linear section ($k_\|<0.05$ \AA$^{-1}$)
differs considerably from both Bi$_2$Te$_3$ and
rocksalt-SnTe\cite{Eremeev_SnTe} $\bar\Gamma$ Dirac states. As a
consequence, the linear fitting for lower and upper sections of the
bandgap TSS (dot-dashed and dashed lines in Fig.~\ref{fig3}(d))
gives $E/k_\|$ slopes of 2.22 and 1.13, respectively. Between these
linear sections, in the shoulder region at
$0.05<k_\|<0.1$~\AA$^{-1}$, two inflection points exist which limit
almost flat section in the TSS spectrum. Such very low velocity
regime in the bandgap TSS resembles heavy-fermion topological state
that has been proposed theoretically\cite{Alexandrov} and obtained
from magnetothermoelectric transport measurements\cite{Luo} in
topological Kondo insulator (TKI) SmB$_6$. Owing to slightly
negative dispersion of the TSS in the shoulder region (see narrow
yellow stripe in Fig.~\ref{fig3}(d)) by varying the chemical
potential the system will pass from one velocity mode in the lower
linear section to the other velocity mode in the upper linear
section through the three surface Fermi pockets regime at energy
$\approx$50 meV (Fig.~\ref{fig3}(f)).

The detailed spin texture of the bandgap TSS is illustrated in
Figs.~\ref{fig3}(g,h). Fig.~\ref{fig3}(g) shows spin-resolved
constant energy contours (CEC's) for the lower (inner contour) and
upper (outer contour) linear sections of the TSS. Both contours,
apart from the in-plane clockwise spin polarization, also
demonstrate the presence of the out-of-plane spin component, which
is intrinsic feature of the spin-polarized states at hexagonal
surfaces while $S_z$ for the inner contour is extremely small. As
clearly seen for the outer contour, $S_z$ is zero along
$\bar\Gamma$-$\bar{\rm M}$ and reaches a minimum (maximum) value in
$\bar\Gamma$-$\bar{\rm K}$ directions. The dependencies of absolute
values for $S_\|$ and $S_z$ on $k_\|$ along $\bar\Gamma$-$\bar{\rm
K}$ are shown in Fig.~\ref{fig3}(h). The in-plane spin polarization
having maximum near the band crossing point rapidly decreases
approaching the shoulder. At $k_\| > 0.1$ \AA$^{-1}$, i.e. in the
upper linear section, it increases with $k_\|$ until it reaches the
region of the bulk states. In contrast, the out-of-plane spin
polarization increases with $k_\|$ in both linear sections of the
TSS, having a kink at the band shoulder.

\begin{figure}
\includegraphics[width=0.75\columnwidth]{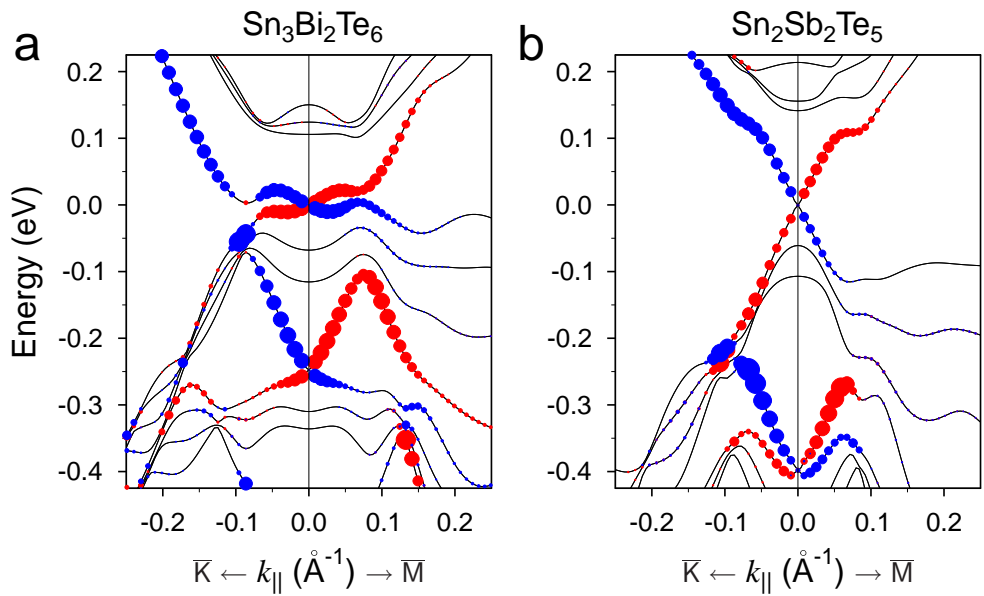}
\caption{(Color online) Spin-resolved surface electronic structure
of Sn$_3$Bi$_2$Te$_6$ (a) and Sn$_2$Sb$_2$Te$_5$ (b).}
 \label{fig4}
\end{figure}

The increase of the SnTe layers in the building block of
(SnTe)$_{n}$(Bi$_2$Te$_3$)$_{m=1}$ compound up to $n=3$ maintains
the two TSSs in the surface spectrum as well as dog-leg behavior in
dispersion of the bandgap surface state (Fig.~\ref{fig4}(a)). The
difference from the Sn$_2$Bi$_2$Te$_5$ case is that the band
shoulders lie closer to the bulk valence band making lower section
of the TSS very short and two times less steep.

Another way to modify the parameters of the bandgap TSS is
substitution of Bi by Sb. In this case SnTe layers are introduced
into Sb$_2$Te$_3$ TI forming (SnTe)$_{n}$(Sb$_2$Te$_3$)$_{m=1}$
compound. As can be seen in Fig.~\ref{fig4}(b) the surface spectrum
of Sn$_2$Sb$_2$Te$_5$ also holds two TSSs with dog-leg dispersion in
the bandgap TSS. However, in contrast to the previous case the band
shoulders arise near the conduction band.

\section{Summary}

In summary, on the basis of {\em ab initio} DFT calculations we have
demonstrated that (SnTe)$_{n=2,3}$(Bi$_2$Te$_3$)$_{m=1}$ compounds
are $\mathbb{Z}_2$ topological insulators in which, owed to
superposition of two pairs of SOI-induced band inversions realized
in SnTe and Bi$_2$Te$_3$ subsystems, two spin-helical topological
surface states arise at the $\bar\Gamma$ point. These states reside
in the band gap and in the local gap of the valence band. The latter
TSS, lying deep below the Fermi level, has a dispersion similar to
that in the Bi$_2$Te$_3$ TI, while the bandgap TSS demonstrates
unusual non-Dirac dog-leg dispersion with two linear sections of
different slope and very low-velocity shoulder section between them.
In Sn$_2$Bi$_2$Te$_5$ the shoulders are located in the middle of the
gap and velocities of carriers in lower and upper linear sections
differ by a factor of two. Thus in contrast to all hitherto known
TIs, where the slope of the spin-helical TSS is linear or varies
smoothly with energy, the bandgap TSS in
(SnTe)$_{n=2,3}$(Bi$_2$Te$_3$)$_{m=1}$ compounds shows abrupt switch
of the velocity of topological carriers with change in the Fermi
surface topology in the narrow energy range (shoulder section) where
spin and charge current should be blocked. We also demonstrated that
similar TSSs exist in (SnTe)$_{n=2}$(Sb$_2$Te$_3$)$_{m=1}$ compound.
This novel finding enhance the understanding on the diversity of
topological surface states and pave a way for the efficient control
of the group velocity with sufficiently large spin current density
by tuning the chemical potential.

Further tuning of the bandgap TSS parameters can be realized by
using (Bi$_{1-x}$Sb$_x$)$_2$Te$_3$\cite{JZhang_NatComm,Ren_PRB2011}
or Bi$_2$(Te$_{1-x}$Se$_x$)$_3$\cite{Ren_PRB2011,Shikin} compounds
as basic TIs for construction of
(SnTe)$_{n}$((Bi-Sb)$_2$(Te-Se)$_3$)$_{m=1}$ systems.


\end{document}